\documentclass[12pt]{article}
\usepackage[top=0.7in,bottom=1in,left=1.1in,right=1.1in,includehead]{geometry}
\usepackage{epsfig,amsmath,amsfonts,verbatim}
\usepackage{lmodern}
\usepackage[T1]{fontenc}
\usepackage{mathtext,marvosym,textcomp}
\usepackage{indentfirst}
\usepackage{epsfig,amsmath,amsfonts,amssymb}
\usepackage{mathtools}
\usepackage[usenames,dvipsnames,svgnames,table]{xcolor}
\usepackage{tikz}
\usetikzlibrary{arrows}
\usetikzlibrary{shapes.misc}
\usetikzlibrary{shapes,positioning,matrix,arrows}
\tikzstyle{every picture}+=[remember picture]
\tikzstyle{na} = [baseline=-.5ex]
\tikzstyle{format} = [rectangle,
                      thick,
                      minimum size=2cm,
                      draw=red!00!black!50,
                      top color=white,
                      bottom color=red!00!black!00,
                      text centered,
                      font=\fontsize{10}{10}\selectfont,
                      on grid]
\tikzstyle{format1} = [rectangle,
                      thick,
                      dashed,
                      minimum size=2cm,
                      draw=red!00!black!50,
                      top color=white,
                      bottom color=red!00!black!00,                     
                      text centered,
                      font=\fontsize{10}{10}\selectfont,
                      on grid]
\tikzstyle{format2} = [font=\fontsize{10}{10}\selectfont,
                      on grid]
\tikzset{cross/.style={cross out, draw=black, minimum size=2*(#1-\pgflinewidth), inner sep=0pt, outer sep=0pt},
cross/.default={5pt}}
 \usepackage[ normalem]{ulem}
 \usepackage{tocloft}
 
 \usepackage[debug,pageanchor=false]{hyperref}
\hypersetup{colorlinks=true,linktocpage,breaklinks,
            urlcolor=blue,
            linkcolor=red,
            citecolor=blue
            }

\numberwithin{equation}{section}

\def\a{\alpha} \def\b{\beta} \def\g{\gamma} \def\d{\delta} \def\e{\epsilon}
  \def\h{\eta} \def\q{\theta}
   \def\l{\lambda} \def\m{\mu}
\def\n{\nu} \def\x{\xi} \def\p{\pi}  \def\r{\rho}
 \def\s{\sigma} \def\t{\tau}  \def\f{\varphi}
\def\ff{\phi}   \def\w{\omega}

\def\G{\Gamma}

   \def\L{\Lambda} 
   
 \def\S{\Sigma}  
\def\F{\Phi}   

\def\ba{\bar{a}}\def\bb{{\bar{b}}}\def\bc{\bar{c}}

 \def\bz{\bar{z}}

\def\bB{\bar{B}}

\def\fr{\frac}  \def\dt{\partial}

\def\gf{\mathfrak{f}}

\def\ph{\phantom}
\def\mc{\mathcal}
\def\mH{\mathcal{H}}

\def\mH{\mathcal{H}}

\def\tx{\tilde{x}}
\def\ty{\tilde{y}}
\def\tz{\tilde{z}}

\def\tdt{\tilde{\partial}}

\def\ww{\wedge}

\def\nn{\nonumber}

\def\hr{\hat{\rho}}

\def\XX{\mathbb{X}}

\def\SS{\mathbb{S}}

\newcommand\bqa {\begin{eqnarray}}
\newcommand\eqa {\end{eqnarray}}

\newcommand{\bear}{\begin{array}}
\newcommand{\enar}{\end{array}}

\newcommand{\br}[2]{\bar{#1}\bar{#2}}

\def\beq{\begin{equation}}
\def\eeq{\end{equation}}
\def\bea{\begin{eqnarray}}
\def\eea{\end{eqnarray}}

\def\F{{\mathcal{F}}}

\def\DD{{\mathcal{D}}}

\begin{document}
\renewcommand{\contentsname}{}
\renewcommand{\refname}{\begin{center}References\end{center}}
\renewcommand{\abstractname}{\begin{center}\footnotesize{\bf Abstract}\end{center}} 
 \renewcommand{\cftdot}{}

\begin{titlepage}
\ph{preprint}

\vfill

\begin{center}
\baselineskip=16pt
   {\large \bf  Non-geometric branes are DFT monopoles}
   \vskip 2cm
    Ilya Bakhmatov$^\dagger$\footnote{\tt ivbahmatov@kpfu.ru}, Axel Kleinschmidt$^{\star,\diamond}$\footnote{\tt axel.kleinschmidt@aei.mpg.de}, Edvard T. Musaev$^{\star\dagger}$\footnote{\tt edvard.musaev@aei.mpg.de}
       \vskip .6cm
             \begin{small}
                          {\it $^\dagger$Kazan Federal University, Institute of Physics\\
                          General Relativity Department\\
                          Kremlevskaya 16a, 420111, Kazan, Russia\\[0.5cm]                          
                          $^\star$Max-Planck-Institut f\"ur Gravitationsphysik (Albert-Einstein-Institut)\\
                          Am M\"uhlenberg 1, DE-14476 Potsdam, Germany \\[0.5cm]
                          $^\diamond$International Solvay Institutes\\
                          Campus Plaine C.P. 231, Boulevard du Triomphe, 1050 Bruxelles, Belgium}\\  
\end{small}
\end{center}

\vfill 
\begin{center} 
\textbf{Abstract}
\end{center} 
The double field theory monopole solution by Berman and Rudolph is shown to reproduce non-geometric backgrounds with non-vanishing Q- and R-flux upon an appropriate choice of physical and dual coordinates. The obtained backgrounds depend non-trivially on dual coordinates and have only trivial monodromies. Upon smearing the solutions along the dual coordinates one reproduces the known $5^2_2$ solution for the $Q$-brane and co-dimension $1$ solution for the $R$-brane. The T-duality invariant magnetic charge is explicitly calculated for all these backgrounds and is found to be equal to the magnetic charge of (unsmeared)  NS5-brane.
\vfill
\setcounter{footnote}{0}
\end{titlepage}

\tableofcontents

\setcounter{page}{2}
\section{Introduction}

Double Field Theory (DFT) has been developed in a number of papers \cite{Hohm:2010jy,Hohm:2010pp,Hull:2009mi,Siegel:1993th} (for reviews see \cite{Berman:2013eva,Hohm:2013bwa,Aldazabal:2013sca}) as a T-duality covariant reformulation of  Type II supergravity. It is a general relativity-like theory with the local diffeomorphism symmetries generated by the T-duality group $O(d,d)$ rather than $GL(d)$. To realise this symmetry one doubles the dimension of the space on which the theory lives by adding new coordinates corresponding to the winding modes of strings:
\begin{equation}
x^\mu \to X^M = (x^\mu, \tilde{x}_\mu) \quad\quad (\mu=1,\ldots,d).
\end{equation} 
Consistency of the algebra of local transformations, given by generalised Lie derivatives \cite{Hull:2009zb,Berman:2012vc}, requires a special constraint called (strong) \textit{section condition}
\begin{equation}
\label{eq:SC}
(\partial_M A(X) )\eta^{MN}  (\partial_N B(X) )= 0 \quad\quad \textrm{for any fields $A(X)$, $B(X)$}
\end{equation}
in terms of the $O(d,d)$ invariant metric $\eta_{MN}\equiv\begin{psmallmatrix}0&1\\1&0\end{psmallmatrix}$. The constraint effectively reduces the number of coordinates leaving only those understood as the physical ones. One possible solution of the section condition is to drop all dependence on the `winding'  coordinates $\tilde{x}_\mu$ and the theory then reproduces conventional supergravity. As a consequence of the T-duality covariance of the theory and the section condition itself, one is allowed to choose the physical subspace of the doubled space in multiple ways, that corresponds to choosing a different T-duality frame.

From the point of view of DFT the space-time metric and Kalb--Ramond field are components of the so-called \textit{generalised metric} $\mH_{MN}$, that is an element of the coset space
\begin{equation}
\mH_{MN} \in \fr{O(d,d)}{O(d)\times O(d)},
\end{equation}
and $\mH_{MN}$ can be understood as a metric on the doubled space. The dilaton $\ff$ together with determinant $g$ of the space-time metric forms an $O(d,d)$ scalar $d=\ff-1/4 \log g$. Components of the generalised metric corresponding to different choices of the physical subspace (determined by a solution to the section condition) are related to each other by Buscher rules~\cite{Buscher:1987qj} thus realising the notion of T-duality transformation in Double Field Theory.

One of the most important applications of the DFT construction is the analysis of non-geometric backgrounds. These are configurations of the supergravity fields defined locally on patches of space-time that are glued together by T-duality transformations \cite{Hull:2004in,Dabholkar:2005ve,Hull:2006qs,Hull:2009sg}. Despite being consistent backgrounds for string theory, they look very exotic from the point of view of supergravity. Although it is not completely clear how to define such configurations within $10$-dimensional supergravity, the result of its compactification on non-geometric backgrounds can be consistently described by non-geometric fluxes and gauged supergravities \cite{Shelton:2005cf}. On the level of conventional supergravity non-geometric fluxes cannot be defined as combinations of fields and their derivatives descending from the $10$- dimensional theory. However, one should note the approach of $\b$-supergravity which is formulated in terms of the metric, dilaton and a bivector field $\b\in\wedge^2TM$, whose derivative is related to the Q-flux \cite{Andriot:2012an,Andriot:2012wx,Andriot:2014uda}. In \cite{Sakatani:2014hba} it has been shown how exotic backgrounds appear as solutions of $\beta$-supergravity written as a ten-dimensional action for non-geometric fluxes.

In contrast, from the point of view of DFT all the fluxes are just components of generalised torsion defined as an $O(d,d)$-covariant bracket of generalised vielbeins $\mH_{MN}=E_M^AE_N^A\mH_{AB}$
\begin{equation}
[E^A,E^B]_C=\F^{AB}{}_CE^C.
\end{equation}
Explicit expressions for the generalised vielbein and components of the generalised torsion can be found in Appendix \ref{torsion}. Considering Scherk--Schwarz reductions of generalised geometry for T- or U-duality groups one recovers half-maximal \cite{Grana:2012rr,Aldazabal:2011nj} and maximal \cite{Musaev:2013rq,Berman:2012uy,Baron:2014yua} gauged supergravities, where the vielbein, now understood as a Scherk--Schwarz twist matrix, is allowed to break the section condition as long as the gauge algebra itself is consistent \cite{Dibitetto:2012rk}. 

In contrast, to construct a $10$-dimensional solution one must satisfy the section condition~\eqref{eq:SC}, however, one is still allowed to keep dependence on the `winding' coordinates. As long as the fields do not depend on mutually dual coordinates simultaneously this respects the section condition. Moreover, from the expressions for the generalised flux one concludes that dual coordinates play crucial role in the definition of non-geometric fluxes, as they are proportional to derivatives of fields along dual coordinates. Hence, one faces the problem of constructing an object, analogous to NS5- or D-branes, sourcing non-geometric fluxes, that may be expected to depend on dual coordinates.

It is important to note here, that although dependence of a background on dual coordinates may look rather exotic from the supergravity point of view, this is not a novel situation in non-linear sigma model \cite{Harvey:2005ab,Jensen:2011jna}. Indeed, in the work \cite{Jensen:2011jna} the background of a Kaluza--Klein-monopole was considered, that is a solution of EOM's of supergravity compactified on a circle $\tz \sim \tz+2\p R_{\tz}$ with harmonic function
\begin{equation}
H(x^i)=1+\fr{h}{\big((x^1)^2+(x^2)^2+(x^3)^2\big)^{\fr12}},
\end{equation}
where $\{x^i\}$ are transverse space coordinates. This is a background with so-called non-zero geometric $\t$-flux dual to NS5-brane smeared along $z$. To recover the version of the KK-monopole localised along the $\tz$ direction one considers worldsheet instanton corrections to the action of non-linear sigma model. These were shown to modify the background precisely in such a way, that the corresponding harmonic function becomes unsmeared and now depends on a new direction $z$ (note the change of power in denominator)
\begin{equation}
H'(x^i,z)=1+\fr{h'}{z^2+(x^1)^2+(x^2)^2+(x^3)^2},
\end{equation}
where $z$ and $\tz$ have the meaning of mutually dual coordinates. Moreover, although the coordinate $\tz$ is not an isometry direction anymore, by applying naive Buscher transformation along this direction one recovers the background of NS5-brane with transverse coordinates $\{z,x^i\}$ and with the harmonic function given by $H'(z,x^i)$. One should note, that such a transformation is an allowed transformation in the framework of DFT, while it is not a symmetry of supergravity solutions. This is the intuitive reason for the dual coordinates to appear in this context.

These ideas were adopted in the papers \cite{Berkeley:2014nza,Berman:2014jsa} where it was shown that the known extended solutions of supergravity equations of motion are just plane waves or Taub--NUT-like solutions from the DFT point of view. In this paper we are interested in the latter, which are referred to as DFT-monopoles. These are solutions of the DFT equations of motion following from the $O(d,d)$ invariant action \cite{Hull:2009mi} and reproducing the NS5-brane and localised KK-monopole solutions upon a choice of physical coordinates. 

In this paper we show that the same DFT-monopole can be used to describe non-geometric backgrounds with non-vanishing Q- and R-fluxes upon yet another choice of the physical slice. The corresponding harmonic function appears to depend on dual coordinates, one for Q-monopole and two for R-monopole. This leads to non-vanishing non-geometric fluxes according to the DFT definition of generalised flux. These fluxes satisfy generalised Bianchi identities and can be used to define the notion of magnetic charge of the DFT solution, which becomes equal to the magnetic charge of the corresponding NS5-brane (unsmeared). This charge is the same for the whole T-duality orbit $H\to \t \to Q \to R$. 

The non-trivial dependence on dual coordinates can be in principle interpreted as a result of the contribution of worldsheet instantons of the corresponding non-linear sigma model. It has been shown in \cite{Kimura:2013zva} that such instanton corrections to the background of the $5_2^2$-brane of de Boer and Shigemori \cite{deBoer:2012ma} lead precisely to a background localised in a dual coordinate. This background is shown to be a smeared version of our Q-monopole solution. 

The paper is structured as follows. In Section \ref{sugra} we consider NS5-brane and KK-monopole as solutions of conventional supergravity and briefly review how worldsheet instanton corrections change the background and introduce dual coordinates. Section \ref{DFTmon} is devoted to the DFT-monopole solution and its non-geometric avatars, which we refer to as Q- and R-monopoles. In Section \ref{fluxes} we calculate components of the generalised flux and explicitly show that Q- and R-monopole indeed source non-geometric Q- and R-fluxes. Finally, in Section \ref{charges} the notion of magnetic charge and Noether current for a DFT solution are considered and the magnetic charge is shown to be precisely equal to that of NS5-brane for DFT-monopole. Appendix~\ref{conventions} contains our conventions for index ranges and other notation.

\section{Dual coordinates from worldsheet instantons}
\label{sugra}

In this section we briefly explain the idea of worldsheet instanton corrections to the background of H- and KK-monopole. A detailed review of these ideas can be found in \cite{Jensen:2011jna,Kimura:2013zva,Tong:2002rq} and in references therein. 

The NS5-brane is a localised brane-like solution of supergravity equations of motion  in $10$ dimensions sourcing a portion of H-flux of magnetic type. Its smeared version that is a solution of EOM's of supergravity compactified along a circle coordinate, say $x^4=z$, is called H-monopole. This is a 5-dimensional object, which from the point of view of Euclidean transverse $(3+1)$-dimensional space with coordinates $(x^i,z)$ looks like a monopole interacting with field strength $H_{zij}$ of magnetic configuration, with $i,j=1,\ldots,3$ in the  transverse direction. 

The background of the smeared NS5-brane that  we refer to as an H-monopole has the following form
\begin{equation}
\begin{aligned}
ds^2&=ds^2_{056789} + H ds^2_{1234},\\
B&=A\wedge dz,\\
e^{-2(\f-\f_0)}&=H^{-1}.
\end{aligned}
\end{equation}
The harmonic function $H(r)=1+h/r$ is a solution of Laplace equation in 3 dimensions and $r^2=\d_{ij}x^ix^j$. ($H(r)$ should not be confused with the flux sourced by the $B$-field that is also referred to as H-flux.) The one form $A=A_idx^i$ plays the role of a gauge field of a magnetic configuration and is given by
\begin{equation}
2\dt_{[i} A_{j]}=\e_{ijk}\dt_k H
\end{equation}
One is able to calculate the magnetic charge of the H-monopole, that is equal to $Q_H=2\p R_z h$, where $R_z$ is radius of the $z$-circle. This is equal to the magnetic charge of the unsmeared NS5-brane and the radius dependence appears from the smearing procedure.

\begin{table}[http]
\centering
\begin{tabular}{|r|cccc|ccccc|}
\hline
           &        1 &        2 &        3 &        4 &        5 &       6 &      7 &        8 & 9         \\  
 \hline
NS5        & $\cdot$ &$\cdot$ & $\cdot$  & $\cdot$  & $\times$ & $\times$ & $\times$ & $\times$ & $\times$    \\
KKM         & $\cdot$ &$\cdot$  & $\cdot$ & $\odot$  & $\times$ & $\times$ & $\times$ & $\times$ & $\times$  \\
 $5^2_2$   & $\cdot$ &$\cdot$ & $\odot$  & $\odot$& $\times$ & $\times$ & $\times$ & $\times$ & $\times$    \\
 \hline 
\end{tabular}
\caption{\sl Under T-dualities an NS5-brane stretched in directions marked by $\times$ turns into a Kaluza-Klein monopole and a $5_2^2$-brane. Dotted circles denote special cycles along which the T-duality acts, these are compactified. }
\end{table}

Performing T-duality along the compact direction $z$ one obtains the background of KK-monopole
\begin{equation}
\label{KKM}
\begin{aligned}
ds^2&=ds_{056789}^2+Hds_{123}^2+H^{-1}(dx^4+A)^2,\\
B&=0. 
\end{aligned}
\end{equation}
Here, the magnetic gauge potential $A_i$ is the $g_{i4}$ component of the metric in the Kaluza-Klein decomposition, hence the name KK-monopole. This background has non-zero geometric flux $\t^z{}_{ij}$.

To do further T-dualities one has to compactify a coordinate, say $x^3$, in order to introduce an isometry direction and smear the monopole along that direction. The smearing procedure reduces the number of transverse directions to $2$ and the harmonic function becomes logarithmically divergent, requiring a cut-off.\footnote{A discussion involving symmetric arrangements of multiple smeared branes can be found in~\cite{Greene:1989ya,Bergshoeff:2006jj}.} The meaning of the cut-off becomes clear if one turns from smearing to solving the Laplace equation in $2$ dimension, that results in a dimensionful integration constant entering the logarithm:
\begin{equation}
\begin{aligned}
H&=1+\sum_{n\in\mathbb{Z}}\fr{h}{\sqrt{\r^2+\big(x^3-2\p \tilde{R}_3 n\big)^2}}\approx 1+\tilde{h} \log\fr{\m}{\r},\\
\r^2&=(x^1)^2+(x^2)^2,
\end{aligned}
\end{equation}
where $\tilde{h}$ is constructed from $h$ and $\m$. Such a harmonic function implies $A=-\tilde{h}\q dx^3$, where $\q$ is the polar angle in the $(1,2)$-plane. Going around the monopole $\q \to \q+2\p $ requires the following gluing conditions that are just diffeomorphism transformations
\begin{equation}
\label{f_twist}
\begin{aligned}
x^3&\to x^3-2\p \tilde{h} x^4,\\ 
x^4&\to x^4.
\end{aligned}
\end{equation}

Performing T-duality along the isometry direction $x^3$ we then arrive at the following background
\begin{equation}
\label{522}
\begin{aligned}
ds^2&=H(d\r^2+\r^2 d\q^2)+\fr{H}{H^2+\tilde{h}^2 \q^2}ds_{34}^2+ds_{056789}^2,\\
B^{(2)}&=\fr{\tilde{h} \q}{H^2+\tilde{h}^2\q^2}dx^3\wedge dx^4,\\
e^{-2(\f-\f_0)}&=\fr{H}{H^2+\tilde{h}^2\q^2},
\end{aligned}
\end{equation}
which is referred to as $5^2_2$-brane and is non-geometric. Indeed, encircling the cycle $\q$ requires gluing the $(x^3,x^4)$-tori at the points $\q=0$ and $\q=2\p$ by a T-duality transformation, that in terms of the generalised metric reads
\begin{equation}
\mc{H}(\q'=\q+2\p)=\mc{O}^{tr}\mc{H}(\q)\mc{O},
\end{equation}
where the matrix $\mc{O}$ encodes the non-geometric $\b$-transform 
\begin{equation}
\label{beta_transf}
\mc{O}=
\begin{bmatrix}
\bf{1}_2 & 0 \\
\b(\q') & \bf{1}_2
\end{bmatrix}
\end{equation}
with $\b(\q)=\tilde{h}\q\, \dt_3\wedge \dt_4$. This suggests to turn to the $\b$-frame of DFT (see Appendix \ref{torsion}), that gives the following background
\begin{equation}
\label{522_beta}
\begin{aligned}
ds^2&=H(d\r^2+\r^2 d\q^2)+H^{-1}ds_{34}^2+ds_{056789}^2,\\
\b&=\b^{34}\fr{\dt}{\dt x^3}\wedge\fr{\dt}{\dt x^4}.
\end{aligned}
\end{equation}
With such an expression at hand one is able to check, that the $5^2_2$-brane is indeed a source of Q-flux, that in this case is just a derivative of the bivector and has one non-zero component $Q_\q{}^{34}$ (for more details on this see \cite{Hassler:2013wsa,Andriot:2012wx,Andriot:2014uda}).

Note however, that the above solution has a logarithmic harmonic function and is a co-dimension 2 object, which causes certain problems concerning it asymptotic behaviour.  Moreover, it is not clear how to confirm that the background of $ 5^2_2$-brane indeed carries Q-flux in the $B$-frame. DFT suggests that in order to see this one has to add dual coordinates into the game, which can be done by considering instanton corrections.

Indeed, applying T-duality to a Kaluza--Klein monopole (a.k.a. Taub--NUT space) along its $S^1$ isometry direction produces an NS$5$-brane with an additional isometry that is commonly referred to as a `smeared' NS$5$-brane. This means that the NS$5$-brane is not completely localised in its four-dimensional transverse space $\mathbb{R}^3\times \mathbb{S}^1$ but has an additional isometry in the transverse $\mathbb{S}^1$ direction along which its charge is smeared homogeneously. However, one can also consider an NS$5$-brane that is localised in the $\mathbb{S}^1$ direction and ask what the T-dual of this configuration in string theory is. This problem was raised in~\cite{Gauntlett:1992nn,Gregory:1997te} and clarified in~\cite{Tong:2002rq} where it was shown that worldsheet instantons play a crucial role.

The simplest way of producing an NS$5$-brane localised in the $\mathbb{S}^1$ direction is to start with flat $\mathbb{R}^4$ as a transverse space and to consider a periodic arrangement of NS$5$-branes along one of its directions that we call $z$. The harmonic function in this case will simply be~\cite{Gauntlett:1992nn}
\begin{align}
H(x^i,z) &= 1 + \sum_{k=-\infty}^\infty \frac{h}{r^2 + (z+2\pi k)^2}\nn\\
&= 1+ \frac{h}{2r} \frac{\sinh r}{\cosh r- \cos z},
\end{align}
where we have chosen the circle to be of unit radius and $r^2=\sum_{i=1}^3 (x^i)^2$ is the distance squared on $\mathbb{R}^3$. The solution is localised at $z=0$ along the $\mathbb{S}^1$. The Fourier expansion of this periodic function in $z$ yields
\begin{align}
\label{eq:Sinst}
H(x^i,z) &= 1 + \frac{h}{2r} \left( 1+ \sum_{k=1}^\infty e^{-kr +i k z} + \sum_{k=1}^\infty e^{-kr -ik z}\right)
\end{align}
which suggests some instanton correction with instanton action $S_{\textrm{inst}} = k r \pm i k z$ to the smeared NS$5$-brane with harmonic function $H=1 + \frac{h'}{r}$. This observation was made precise by Tong~\cite{Tong:2002rq} where he showed that the two-dimensional gauged linear sigma model underlying the smeared NS$5$-brane (that is T-dual to the Kaluza--Klein monopole) receives worldsheet instanton corrections of precisely the type discussed above.\footnote{The `instanton measure' from the supergravity configuration~\eqref{eq:Sinst} above comes out to be equal to one for all instanton charges $k$. This has not been fully confirmed independently from a worldsheet calculation.} In this way, worldsheet instantons are related to localisation in the $\mathbb{S}^1$ direction of the transverse space of the NS$5$-brane.

In the T-dual picture of the Kaluza--Klein monopole this localisation effect does not occur in the usual `momentum' space but in the dual `winding' space of the string. This point of view was emphasised in~\cite{Harvey:2005ab} and the corresponding double field theory interpretation was given later in~\cite{Jensen:2011jna} where it was shown that the worldsheet instantons in this language naturally provide  an origin of dual coordinates after T-duality of a solution that is localised and not smeared. This strategy was later extended to the $5^2_2$-brane in~\cite{Kimura:2013zva} where the smeared $5^2_2$-brane has co-dimension two and can be obtained by performing a further T-duality on the smeared Kaluza--Klein monopole~\cite{Obers:1998fb,LozanoTellechea:2000mc}, see also~\cite{Englert:2007qb,Kleinschmidt:2011vu} for further discussions of duality orbits of smeared co-dimension two objects.

\section{DFT monopole}
\label{DFTmon}

From the point of view of DFT the backgrounds of KK-monopole and H-monopole are particular cases of the solution presented in  \cite{Berman:2014jsa} and called DFT-monopole. Its generalised metric $\mH_{MN}$ has a Taub-NUT form and can be represented as a formal line element on the full $(10+10)$-dimensional space
\begin{equation}
\begin{aligned}
ds^2_{DFT}&=H(1+H^{-2}A^2)dz^2+H^{-1}d\tz^2+2H^{-1}A_i(dy^id\tz-\d^{ij}d\ty_j dz)\\
          &+H(\d_{ij}+H^{-2}A_iA_j)dy^idy^j+H^{-1}\d^{ij}d\ty_id\ty_j\\
          &+\h_{rs}dx^rdx^s+\h^{rs}d\tx_rd\tx_s,
\end{aligned}
\end{equation}
where the functions $H,A_i$ and the invariant dilation are given by ($i,j=1,2,3$)
\begin{equation}
\label{H}
\begin{aligned}
H(y)&=1+\fr{h}{\sqrt{\d_{ij}y^iy^j}},\\
2\dt_{[i}A_{j]}&=\e_{ijk}\dt_kH,\\
e^{-2d}&=He^{-2\f_0},
\end{aligned}
\end{equation}
and the conventions for indices are collected in Appendix \ref{conventions}. Here $\f_0$ and $h$ are some constants parametrizing the solution with $h$ being related to the  magnetic charge of the solution. To address space-time properties of the DFT solution from the supergravity point of view, one should choose a subset of physical coordinates. In other words, one should agree on which subset of the $10$ coordinates $(z,y^i,x^r,\tz,\ty_i,\tx_r)$ are physical and which are dual, i.e.\ corresponding to the winding modes of strings. 

In general, the coordinates $\XX^M$ on the doubled space can be decomposed as follows
\begin{equation}
\XX^M=(x^z,x^i,x^r,\tx_z,\tx_i,\tx_r)
\end{equation}
and we will stick to the convention that $x^{z,i,r}$ will always denote physical coordinates, while $\tx_{z,i,r}$ will be always the dual ones. However, this still does not tell anything about the duality frame for the solution we are analysing, as one has to identify the parameters $z,y^i, \ldots$ with the coordinates above. Depending on the way this is done, the above solution of DFT yields different solutions of Type II supergravity. For example, the choice $(x^z,x^i)=(z,y^i)$ gives the conventional H-monopole solution of supergravity, while the rule $(x^z,x^i)=(\tz,y^i)$ corresponds to KK-monopole.

To identify the supergravity fields $g_{\m\n}, B_{\m\n}$ (or $\b^{\m\n}$) and $\f$ one considers DFT as a Kaluza--Klein theory and writes the DFT line interval as
\begin{equation}
\label{eq:genM}
ds^2_{DFT}=(g_{\m\n}-B_{\m}{}^\r B_{\r\n})dx^\m dx^\n+2B_\m{}^\n dx^\m d\tx_\n+g^{\m\n}d\tx_\m d\tx_\n.
\end{equation}
Choosing the subset of physical coordinates and comparing the DFT solution with the above ansatz one uniquely identifies the 10-dimensional fields.

\subsection{H- and KK-monopole}

For completeness of the narration let us start with H- and KK-monopole, which are conventional geometric backgrounds of Type II supergravity, and repeat the results of \cite{Berman:2014jsa}. As was discussed in the previous section, the H-monopole is a smeared version of the NS5-brane while the KK-monopole solution is its T-dual along the smearing coordinate. In the framework of DFT these are just two faces of the single DFT monopole solution. Indeed, choosing the physical coordinates to be
\begin{equation}
x^\m=(z,y^i,x^r)
\end{equation}
one obtains the NS5-brane solution smeared along the $z$ direction
\begin{equation}
\begin{aligned}
ds^2&=\h_{rs}dx^rdx^s+H(dz^2+\d_{ij}dy^idy^j),\\
B&=A_idy^i\wedge dz,\\
e^{-2(\f-\f_0)}&=H^{-1}.
\end{aligned}
\end{equation}
In the notation of the classification \cite{deBoer:2012ma} this is the $5^0_2$-brane. As will be shown in Section \ref{fluxes}, this background interacts with H-flux with non-vanishing component being $H_{zij}=\e_{ijk}\dt_k H$.

From the point of view of conventional supergravity smearing is necessary to make connection between the NS5-brane and the KK-monopole ($5^1_2$-brane of \cite{deBoer:2012ma}) via T-duality along the (compact) direction $z$. This procedure allows to reproduce the harmonic function of the solution of compactified theory from the harmonic function of the full solution  without having to solve the equations of motion from the very beginning \cite{Ortin:2015hya}. Physically this is interpreted as putting an infinite number of branes with distance $2\p R_z$ between them and summing all contributions to the harmonic function $H$. Sending $R_z \to 0$ this effectively corresponds to dropping any dependence on $z$.

Now, T-duality along $z$ in the DFT picture corresponds to replacing $x^z$ by its dual $\tx_z$, i.e. to  choosing the following set of coordinates to be physical
\begin{equation}
x^\m=(\tz,y^i,x^r). 
\end{equation}
This gives the background of the KK-monopole solution
\begin{equation}
\begin{aligned}
ds^2&=\h_{rs}dx^rdx^s+H^{-1}(d\tz+A_idy^i)^2+H\d_{ij}dy^i dy^j,\\
B&=0,\\
e^{-2(\f-\f_0)}&=1.
\end{aligned}
\end{equation}

It is important to note, that although we consider only the monopole versions of the corresponding brane configurations (smeared along one direction), the full solution has been presented in \cite{Berman:2014jsa} as well. The authors refer to it as a localised KK-monopole, however we would prefer to call it KK-brane (Q-brane, R-brane), reserving the word ``localised'' for solutions living on compact $x^z$, but with finite $R_z$. The corresponding harmonic function still depends on $x^z$ and contains all Fourier modes in~\eqref{eq:Sinst}, in contrast to the smeared solution, which contains only the zero mode. According to \cite{Tong:2002rq,Jensen:2011jna,Kimura:2013zva}, this is precisely the harmonic function that is recovered by considering instanton corrections. These naturally require a periodic coordinate to contribute to the worldsheet action of the sigma model. This is not necessary in the DFT picture, which naturally reproduces the desired harmonic function upon compactification of a dual coordinate with finite radius.

\subsection{Q-monopole}

The non-geometric $5^2_2$-brane of Shigemori and de Boer is obtained by smearing the $5^1_2$-brane solution presented in the previous section along, say, $y^3$ and performing T-duality along this (now) compact direction. However, we will act in a more direct way and obtain it from the DFT monopole solution by choosing
\begin{equation}
\label{Q-phys}
x^\m=(\tz,y^1,y^2,\ty_3,x^r). 
\end{equation}
After some algebra reading off the components of the fields from the generalised metric~\eqref{eq:genM}, one obtains the following $10$-dimensional background
\begin{equation}
\label{Q}
\begin{aligned}
ds^2&=\h_{rs}dx^rdx^s+\fr{H}{H^2+A_3^2}\Big((d\tz+A_\a dy^\a)^2 + d\ty_3^2\Big)+H\d_{\a\b}dy^\a dy^\b,\\
B&=\fr{A_3}{H^2+A_3^2}(d\tz+A_\a dy^\a) \wedge d\ty_3,\\
e^{-2(\f-\f_0)}&=\fr{H}{H^2+A_3^2},
\end{aligned}
\end{equation}
where $\a,\b=1,2$ label the coordinates $y^{1,2}$. Note that the harmonic function $H$ depends now on 
the winding coordinate $y^3$
\begin{equation}
H=1+\fr{h}{\sqrt{\d_{\a\b}y^\a y^\b+(y^3)^2}}.
\end{equation}
Smearing this harmonic function along $y^3$ with finite radius gives precisely the instanton-corrected harmonic function of \cite{Kimura:2013zva} smeared along $X^9$ ($z$ in our notations). For that one considers an infinite array of Q-monopoles along $y^3$ and writes
\begin{equation}\label{log}
\begin{aligned}
H &= 1 + \sum_{k=-\infty}^\infty \frac{h}{\sqrt{\d_{\a\b}y^\a y^\b+(y^3+2\pi k)^2}}\\
 &=1+h\log \dfrac{\L + \sqrt{\L^2 +\r^2 }}{\r^2}\approx h_0+h\log{\fr{\m}{\r}}
\end{aligned}
\end{equation}
where $\r^2=\d_{\a\b}y^\a y^\b$. Here the divergent sum has been replaced by a divergent integral and the cut-off $\L$ has been introduced. The first expression in the second line diverges as $\L \to \infty$ however it can be rewritten by introducing a bare quantity $h_0$, which also diverges in this limit, and a renormalization scale $\m$ (see \cite{deBoer:2012ma}). Hence, the harmonic function of the Q-monopole recovers the harmonic function of the known $5^2_2$-brane upon smearing. To investigate the metric and the B-field, we switch to polar coordinates on the $(y^1,y^2)$ plane for convenience
\begin{equation}
 \begin{aligned}
  y^1&=\r \cos \q,\\
  y^2&=\r \sin \q.
 \end{aligned}
\end{equation}
With this set up we have the following equations for the vectors $A^\a$ and $A^3$
\begin{equation}
 \begin{aligned}
  \r\dt_\r H&=\dt_\q A_3,\\
  0&=\dt_\r A_\q-\dt_\q A_\r,\\
  0&=\dt_\r A_3.
 \end{aligned}
\end{equation}
The second line above implies that the components $\{A_r,A_\q\}$ are given by just a gauge degree 
of freedom
\begin{equation}
 \begin{aligned}
  A_\r&=\dt_\r \l,\\
  A_\q&=\dt_\q \l,
 \end{aligned}
\end{equation}
with $\l=\l(r,\q)$ being an arbitrary function. The first line fixes the remaining component to be 
$A_3=h \q$. Redefining the coordinate $\tz$ as $\tz \to \tz+\l$ we arrive at the familiar 
background with nontrivial monodromy around $\q$
\begin{equation}
\label{522B}
\begin{aligned}
ds^2&=\h_{rs}dx^rdx^s+HK^{-1}\Big(d\tz^2 + d\ty_3^2\Big)+H\d_{\a\b}dy^\a 
dy^\b,\\
B&=h \q K^{-1}d\tz \wedge d\ty_3,\\
e^{-2(\f-\f_0)}&=HK^{-1},\\
K&=H^2+(h \q)^2.
\end{aligned}
\end{equation}

There is a certain subtlety in understanding this background. In particular, although it is in 
general accepted that this background generates a non-trivial Q-flux, the direct calculation of 
$Q^{mn}{}_{k}$ using the B-frame of DFT gives a vanishing result. On the other hand, performing the 
above steps in the $\b$-frame we arrive at the following background
\begin{equation}
\label{522beta}
\begin{aligned}
ds^2&=\h_{rs}dx^rdx^s+H(d\tz^2 + d\ty_3^2+\d_{\a\b}dy^\a dy^\b),\\
\b&=h \q \dt_{\tz} \wedge \dt_{\ty_3},
\end{aligned}
\end{equation}
which clearly has a non-trivial component of the Q-flux, that is $Q^{z3}{}_{\q}=h$. On the other 
hand, the above background is completely geometric given the gauge transformations of 
$\b$-supergravity
\begin{equation}
 \d\b^{\m\n}=\w^{\m\n}=\mbox{const}
\end{equation}
This suggests that the dropped winding coordinate $y^3$ plays an important role in the
identification of the solution as carrying a portion of Q-flux.

Consider now the full solution not smeared along $y^3$ with the fields depending on the full set of coordinates $\{y^1,y^2,y^3\}$ and the coordinate $y^3$ being understood as a winding mode. In this case it is convenient to turn to cylindrical coordinates as
\begin{equation}
\begin{aligned}
y^1&=\r \cos{\q},\\
y^2&=\r \sin{\q},\\
y^3&=y^3.
\end{aligned}
\end{equation}
The equations defining the gauge field $A$ then take the following form
\begin{equation}
\begin{aligned}
\r\dt_3 H&=\dt_\r A_\q-\dt_\q A_\r\\
0&=\dt_3 A_\r-\dt_\r A_3,\\
\r\dt_\r H&=\dt_\q A_3-\dt_3 A_\q.
\end{aligned}
\end{equation}
As usual this is complemented by the conditions div$A=0$ and $\triangle A=0$. The solution is of the Taub-NUT type with the only non-vanishing component being
\begin{equation}
A_\q=h\Bigg(1-\fr{y^3}{\sqrt{\r^2+(y^3)^2}}\Bigg).
\end{equation}
The most interesting issue here is, that the resulting background is purely of the metric type, i.e. the $B$- or the $\b$-field vanish in either frame. In addition, in both frames the metric is given by the same expression
\begin{equation}
ds^2=H^{-1}\Big[\big(d\tz+A_\q d\q\big)^2 + d\ty_3^2\Big] + H\Big( d\r^2+\r^2 d\q^2\Big).
\end{equation}
The solution is localised along the winding coordinate $y^3$, which is of no surprise given the above dicsussion of worldsheet instanton corrections. However, one should investigate the additional information provided by the fact, that $y^3$ is not periodic.

As the original Kaluza--Klein monopole, the above solution suffers from the Taub-NUT singularity, which is a pure coordinate singularity in the case when $\tz$ is a compact coordinate. This is actually the case, as we have started from the non-localised KK-monopole solution, meaning the coordinate $\tz$ is compactified with small radius.

It is tempting to claim, that the Q-brane is just an analogue of the conventional KK-monopole solution, but with one coordinate replaced by its winding counterpart. This is also expected, as these are T-dual to each other, however the above shows that by a direct computation from DFT. For this reason and in analogy with the H- and KK-monopole solutions we will call this solution Q-monopole. 

Note that the Q-monopole solution is completely geometric in the aspect of the monodromy property, i.e. going around the monopole in the $\{y^1,y^2\}$-plane by shifting $\q\to \q+2\p$ does not change the solution. Although, there are still signs of non-geometry represented by the dependence on the winding coordinate $y^3$ which lead to non-vanishing Q-flux as we show further.

\subsection{R-monopole}

Backgrounds with R-flux are the most subtle in the T-duality orbit in question as they correspond to  codimension-1 objects that are obtained by a T-duality action along a non-isometry direction. However, from the point of view of DFT a T-duality transformation is just an $O(d,d)$ rotation, that replaces a coordinate by its dual. Hence, one may consistently consider such backgrounds by studying the following choice of physical coordinates
\begin{equation}
x^\m=(\tz,y^1,\ty_2,\ty_3,x^r). 
\end{equation}
Due to the reasons explained below this background should be considered in the $\b$-frame, that gives the following 
\begin{equation}
\begin{aligned}
ds^2&=\h_{rs}dx^rdx^s+H^{-1}\Big((d\tz+A_1 dy^1)^2 + d\ty_\a^2\Big)+H(dy^1)^2,\\
\b&=A_\a \dt_{\ty_\a}\wedge \dt_{\tz},\\
e^{-2(\f-\f_0)}&=1,
\end{aligned}
\end{equation}
where now $\a=2,3$. As in the previous case the convenient choice of the coordinate frame is the cylindrical coordinates, however now the distinguished coordinate is $y^1$ and the rules read
\begin{equation}
\begin{aligned}
y^2&=\r \cos{\q},\\
y^3&=\r \sin{\q},\\
y^1&=y^1.
\end{aligned}
\end{equation}
 Taking again the Taub-NUT solution and adapting it to the chosen coordinate frame we have
\begin{equation}
\begin{aligned}
A_\q&=h\Bigg(1-\fr{y^1}{\sqrt{\hat{\r}^2+(y^1)^2}}\Bigg),\\
\hat{\r}^2&=(y^2)^2+(y^3)^2.
\end{aligned}
\end{equation}
The solution now depends on two winding coordinates and the background becomes
\begin{equation}
\begin{aligned}
ds^2&=\h_{rs}dx^rdx^s+H^{-1}\Big(d\tz^2 + d\hat{\r}^2+\hat{\r}^2d\q^2\Big)+H(dy^1)^2,\\
\b^{\q z}&=A_\q, \\
e^{-2(\f-\f_0)}&=1.
\end{aligned}
\end{equation}
The solution looks like very similar to that of the NS5-brane, but with one distinguished transverse direction. This can be interpreted by the R-brane being a co-dimension 1 object. In analogy to the NS5-brane one expects the above solution to have only $R^{\q z 1}$ flux.

One may wonder what happens if the above background is written in the B-frame. Given the definition of the generalised flux the obvious answer is that the background will no longer have R-flux. This is of no surprise, as for example the H-monopole background written in the $\b$-frame does not carry H-flux anymore. 

Here we observe an interesting symmetry of the T-duality orbit $H\to \t \to Q \to R$. It tells us that the backgrounds at the H-node and R-node require a certain supergravity frame to be consistently written down. However, the $\t$- and Q-nodes are purely metric backgrounds which do not depend on the frame chosen. Hence, we may speculate that H-monopole and R-monopole are backgrounds with magnetic configurations of the $B$-field or $\b$-field respectively. However, it is unclear what this means for the bivector, as it does not have the required gauge transformations for its component $\b^{zi}$ to be interpreted as a gauge potential. 

The same applies to KK-monopole, that is understood as an object interacting with the magnetic gauge field coming from the metric $A_i=g_{iz}$. It is tempting to interpret the Q-monopole in a similar fashion but using the inverse (dual in a sense) metric $g^{zi}$, however we could not go much further in that direction.

\section{Fluxes and Bianchi identities}
\label{fluxes}

The T-duality orbit that relates the backgrounds considered above consists of four points represented by the following fluxes
\begin{equation}
\begin{aligned}
H_{mnk} && \longleftrightarrow && \t^m{}_{nk} && \longleftrightarrow && Q_m{}^{nk} && \longleftrightarrow && R^{mnk}.
\end{aligned}
\end{equation}
In what follows we assume that the fluxes live in a four-dimensional space relevant for our discussion, however all the arguments below are valid for any dimensions.

The $H$- and R-fluxes belong to the irreducible representation $\bf 4$ of $SO(4)$, while the fluxes $\t^{\ba}{}_{\bb\bc}$ and $Q_{\ba}{}^{\bb\bc}$ can be in principle decomposed as
\begin{equation}
\begin{aligned}
\t^{\ba}{}_{\bb\bc}: && \bf 4 \otimes \bar 6 \longrightarrow \bar 4 \oplus 20\\
Q_{\ba}{}^{\bb\bc}: && \bf \bar 4 \otimes 6 \longrightarrow      4 \oplus \overline{20}.
\end{aligned}
\end{equation}
The trace part of the geometric flux is usually restricted to be zero, when considering compact Calabi--Yau manifolds. However, when understood as full 10-dimensional quantities, trace parts of both geometric and Q-flux are not necessarily zero.

From the DFT point of view these fluxes are components of the generalised torsion $\F^A{}_{BC}$ defined by (see Appendix \ref{torsion})
\begin{equation}
\begin{aligned}
[E_B,E_C]^M_C&=\F^{A}{}_{BC} E^M_A,\\
\F^{A}{}_{BC}&=2E^{A}_M E_{[B}^N\dt_N E_{C]}^M-E^{A}_M\h^{MN}\h_{KL}\dt_N E_{[\bB}^K E_{C]}^L.
\end{aligned}
\end{equation}
In addition to that one has the flux $\F_A=\dt_M E^M_A+2E^M_A\dt_M d$, which vanishes for ordinary compactification scenarios.

For completeness of the picture let us start with fluxes of the geometric backgrounds of H and KK monopoles. For the H-monopole solution one has both the H-flux and the trace part of the $f$-flux, their non-vanishing components being
\begin{equation}
\begin{aligned}
\mbox{H-monopole:}&& \mH_{\bar{z}\ba\bb} = 2e^z{}_{\bar{z}}e^k{}_{\ba}e^l{}_{\bb}\dt_{[k} A_{l]}, && \F^{\ba}{}_{\bb\bc} = -\d^{\ba}{}_{[\bb}\gf_{\bc]},  && \F_{\ba} = \frac32\, \gf_{\ba},
\end{aligned}
\end{equation}
where $\gf_{\ba}= H^{-1}\dt_{\ba} H$. 
As expected, the H-monopole solution interacts with the field strength $\mH_{zkl}$ of a magnetic configuration, whose gauge potential is given by the  Kalb-Ramond field $B_{zk}$. Hence the name H-monopole.

For the KK-monopole we have the following non-vanishing fluxes
\begin{equation}
\begin{aligned}
\mbox{KK-monopole:}&& \F^{\ba}{}_{\bb\bc} = 2e^{\ba}{}_{z}e^k{}_{\bb}e^l{}_{\bc}\dt_{[k} A_{l]} - \fr13\d^{\ba}{}_{[\bb}\gf_{\bc]}, && \F_{\ba} = -\frac32 \gf_{\ba}.
\end{aligned}
\end{equation}
According to its name, the KK-monopole solution interacts with the field strength $\F^{z}{}_{ij}$ of the same magnetic configuration, whose gauge potential is given by the component $A_i=g_{zi}$ of the metric.

Although the Q-monopole solution does not contain non-trivial gauge fields, dependence on the winding coordinate $y^3$ makes the Q-flux non-zero. The following non-vanishing components
\begin{equation}
\begin{aligned}
                   && Q_{\bar{\a}}{}^{\bar{3}\bz}&=\e_{\bar{\a}\bar{\b}}H^{-1}\dt_{\bar{\b}}H,  
                       & \F^{\bar{\a}}{}_{\br{1}{2}}&=\fr12 \e^{\br{\a}{\b}}H^{-1}\dt_{\bar{\b}}\\
                   && Q_{\bar{\a}}{}^{\br{\b}{3}}&=-\fr12 \d^{\bar{\b}}{}_{\bar{\a}}H^{-\fr32}\dt_3 H,
                       & \F^{\bar{3}}{}_{\br{\a}{3}}&=\fr12 H^{-1}\dt_{\bar{\a}}H,\\
\mbox{Q-monopole:} && Q_{\bz}{}^{\br{z}{3}}      &=\fr12 H^{-\fr32} \dt_3 H,
                       & \F^{\bz}{}_{\br{\a}{\b}}&=-\e_{\br{\a}{\b}}H^{-\fr32}\dt_3 H,\\
                   && & 
                   		& \F^{\bz}{}_{\br{\a}{z}}&=\fr12 H^{-1}\dt_{\bar{\a}}H  ,\\
                   && \F^{\bar{3}} &=\fr32 H^{-3/2}\dt_3 H 
                   		& \F_{\bar{\a}}&=\fr32H^{-1}\dt_{\bar{\a}}H.
\end{aligned}
\end{equation}
where one should note that $\dt_3$ is the derivative along a winding mode coordinate.  These are the most general expressions for the fluxes of the Q-monopole background which do not depend on the coordinates adopted to solve the equations for $A_m$. One immediately notices, that the flux components $Q_{\bar{\a}}{}^{\bar{3}\bz}$ are the same as in the case of the $5^2_2$ background up to the explicit form of  the harmonic function. After smearing along the winding coordinate $y^3$ this gives in curved indices the only component $Q_{\q}{}^{z3}=h$.

Let us finally turn to the R-monopole background that turns on the R-flux. Together with other non-vanishing flux components we have
\begin{equation}
\begin{aligned}
                  && R^{\bar{z}\br{\a}{\b}}&=\e^{\br{\a}{\b}}H^{-\fr32} \dt_1 H, 
                     & Q_{\bar{z}}{}^{\br{z}{\a}}&=Q_{\bar{1}}{}^{\br{1}{\a}}=-\fr12 H^{-1}\tdt^{\bar{\a}}H,\\
\mbox{R-monopole:} && \F_{\ba}{}^{\bb\bc}&=\d_1{}^{[\bb}\d_{\ba}{}^{\bc]}H^{-\fr32}\dt_{1}H,
                     & Q_{\bar{\a}}{}^{\br{\b}{\g}}&=\fr12\e^{\br{\b}{\g}}\e_{\br{\a}{\d}} H^{-1}\tdt^{\bar{\d}}H,\\
                  &&               &
                     &  Q_{\bar{1}}{}^{\br{z}{\a}}&=\e^{\br{\a}{\b}} H^{-1}\tdt^{\bar{\b}}H  \\
                  && \F^{\bar{\a}} &=\fr32 H^{-1}\tdt_{\bar{\a}} H 
                                     		& \F_{\bar{1}}&=\fr32H^{-\fr32}\dt_{1}H.   
\end{aligned}
\end{equation}
where $\a,\b$ take the values 2 and 3. Hence, there is a non-trivial Q-flux and R-flux sourced by the background, while the geometric flux $\F_{\ba}{}^{\bb\bc}$ only has a trace part. Again by formally smearing the solution along the winding coordinates $\{y^2,y^3\}$ the Q-flux components vanish and one ends up with only the R-flux and 
the geometric flux.

To make things more manifest the R-flux can be rewritten as
\begin{equation}
R^{\bar{z}\br{\a}{\b}}=-2\,0e^{\bz}{}_{z}e^{\bar{\a}}{}_{\a}e^{\bar{\b}}{}_{\b}\tdt^{[\a} \b^{\b]z}.
\end{equation}
This is consistent with the interpretation of the R-flux as a field strength for the $\b$-field. Hence, the R-monopole interacts with a ``magnetic'' part of the $R^{zij}$ component of this field strength. The corresponding gauge transformation is given by $\d \b^{\m\n}=2\tdt^{[\m}\l^{\n]}$.

All these flux components satisfy the generalised Bianchi identities of \cite{Geissbuhler:2013uka} which are explicitly written out in the Appendix \ref{bianchi}.

\section{Charges and currents}
\label{charges}

\subsection{Magnetic charge}
\label{charge}

Before proceeding with the DFT construction let us look at how the notion of magnetic  and electric charges is defined in conventional electromagnetism. We write
\begin{equation}
\begin{aligned}
4\p q&=\int \dt_i E^idV=\int_\S E^i d\S_i=\int_\S F_{0i}d\S^{0i}=\int F_{\m\n}d\S^{\m\n}=\int _\S *F,\\
4\p \m&=\int \dt_i B^idV=\int_\S B^i d\S_i=\int_\S F_{kl}\e^{0ikl}d\S_{0i}=\int F_{\m\n}dx^\m\ww dx^\n=\int _\S F,
\end{aligned}
\end{equation}
where $\S$ is a spacelike surface surrounding the charge, say a 2-sphere, $dx^\m \wedge dx^\n = \e^{\m\n\r\s} d\S_{\r\s}$, and $F=dA$ is the gauge field strength (flux). Hence, the magnetic charge can be defined as an integral of the flux along a 2-cycle.

Motivated by this we adopt the definition of the DFT magnetic charge from \cite{Blair:2015eba} and write
\begin{equation}\label{magn-charge}
4\p \m=\int_\S \F_{MNK}d\XX^M \wedge d\XX^N \wedge d\XX^K,
\end{equation}
with appropriately chosen 3-cycle $\S$ which is a three-dimensional surface surrounding a monopole considered as a point in the 4-dimensional doubled space parametrised by $\{x^z,x^i\}$. Here, $x^\m$ should be properly identified with the coordinates of the doubled space according to a solution of the section condition.

This immediately tells us that the surface cannot be non-trivially defined by varying two mutually dual coordinates, say $z$ and $\tz$, as this clearly breaks the section condition. This naturally removes the components of generalised flux of the type $\F^a{}_{aM}$ (no sum). The remaining components themselves suggest which surface to choose to get a non-vanishing result (see Table  \ref{sigma}). Since all the solutions we consider here are smeared along $x^z$ the topology of the 3-cycle is restricted to be of the type $\SS^2\times \SS^1$ instead of a 3-sphere, which one should expect in the localised case. It is important to note here, that although the surface spans  different coordinates for different solutions, the $\SS^2$ part is always parametrised by the equation
\begin{equation}
(y^1)^2+(y^2)^2+(y^3)^2=R^2=\mathrm{const}.
\end{equation}
This is due to the identification of (some of) the coordinates $\{y^i\}$ with either $\{x^i\}$ or $\{\tx_i\}$.

\begin{table}[http]
\centering
\begin{tabular}{|c|cccccccc|}
\hline
$\S$ & $x^1$    &$ x^2 $   & $x^3$    & $x^z $    &$ \tx_1$  & $\tx_2$ & $\tx_3$ & $\tx_z $  \\  
 \hline
H   & $\times$ & $\times$ & $\times$ & $\bullet$ &          &         &         &           \\
KK  & $\times$ & $\times$ & $\times$ &           &          &         &         & $\bullet$ \\
Q   & $\times$ & $\times$ &          &           &          &         & $\times$& $\bullet$ \\
R   & $\times$ &          &          &           &          & $\times$& $\times$& $\bullet$  \\
 \hline 
\end{tabular}
\caption{\sl The 3-cycle $\S \equiv \SS^2\times \SS^1$ is a product of a 1-circle and a 2-sphere. Here the bullet $\bullet$  denotes the direction of the 1-circle, while the crosses $\times$ denote the directions in which the 2-sphere lives.}
\label{sigma}
\end{table}

With all this in hands let us turn to explicit computations and start with the H-monopole solution which has the only relevant flux component $\F_{zij}=2\dt_{[i}A_{j]}=\e_{ijk}\dt_k H$. Hence, we write
\begin{equation}
\begin{aligned}
4\p\m^H&=\int \F_{zij}dx^z\ww dy^i\ww dy^j=\int \F_{zij}dz\ww dy^i\ww dy^j=2\p R_z \int  \sin \q d\q d\f R^2 \dt_r H(r)\Big|_{r=R}\\
&=8\p^2 R_z h=4\p Q,
\end{aligned}
\end{equation}
where $Q=2\p R_z h$ is the charge of the \textit{unsmeared} NS5 brane defined by the harmonic function
\begin{equation}
H_{unsm}=1+\fr{Q}{r^2}.
\end{equation}
Note that all other components of the generalised flux do not contribute to a magnetic charge defined this way.

The same calculation for the KK-monopole with the relevant flux component $\F^z{}_{ij}=\e_{ijk}\dt_k H$ gives
\begin{equation}
4\p\m^{KK}=\int \F^z{}_{ij}d\tx_z\ww dy^i\ww dy^j=\int \F_{zij}dz\ww dy^i\ww dy^j=8\p^2 R_z h.
\end{equation}
The important change here is that now one integrates over $\tx_z$ which for the chosen section condition frame is still $z$, and hence the integral itself does not change. A similar effect will take place for non-geometric Q- and R-monopoles. The 3-dimensional surface $\S$ is now partially stretched in the dual space, i.e. along the coordinate $\tx_z$. However, as expected, the magnetic charge for H- and KK-monopole is the same as these belong to the same T-duality orbit, while the definition of $\m$ is T-duality invariant.

The relevant flux components (in curved indices) for the Q-monopole are given by
\begin{equation}
\begin{aligned}
\F^z{}_{12}&=-\dt_3 H-A_1 H^{-1}\dt_2 H+A_2 H^{-1}A_2 \dt_1 H, \\
\F_1{}^{3z}&=\dt_2 H-A_1 H^{-1}\dt_3 H,\\
\F_2{}^{3z}&=-\dt_1 H-A_2 H^{-1}\dt_3 H.
\end{aligned}
\end{equation}
The magnetic charge then reads
\begin{equation}
\begin{aligned}
4\p\m^Q=&\int \F^z{}_{12}d\tx_z\ww dx^1 \ww dx^2+ \int \F_1{}^{3z} dx^1\ww d\tx_3\ww d \tx_z+\int \F_2{}^{3z} dx^2\ww d\tx_3\ww d \tx_z\\
=&\int \F^z{}_{12}dz\ww dy^1 \ww dy^2+ \int \F_1{}^{3z} dy^1\ww dy^3\ww d z+\int \F_2{}^{3z} dy^2\ww dy^3\ww d z\\
=&\int \Big[ -\r \dt_3 H dz\ww d\r \ww d\q +\r\big(-A_1 \dt_2 H+A_2\dt_1 H\big)dz\ww d\r\ww d\q \\ 
 &+\big(\dt_2 H dy^1-\dt_1 H dy^2\big)\ww dy^3 \ww dz -\big(A_1 dy^1+A_2 dy^2\big)H^{-1}\dt_3 H \ww dy^3 \ww dz \Big]\\
=&\int  \big(\r \dt_3 H  d\r-\r\dt_\r H dy^3\big)dz d\q-\r A_\q\big(\dt_\r H d\r+\dt_3 H dy^3\big)dz d\q\\
=&-\int \big(y^3 \dt_3 H+\r \dt_\r H\big)dzdy^3d\q=8\p^2 R_z h,
\end{aligned}
\end{equation}
where we have used that $\dt_\r H d\r+\dt_3 H dy^3=dH=0$  and the constraint $\r d\r+y^3dy^3=0$ on the integration surface.

As before, due to a different identification of dual and physical coordinates, the integral is reduced to an integral in the $\{z,y^1,y^2,y^3\}$ space. However, now the 2-sphere $\SS^2$ is partially lying in the dual space. This result is to be expected as will be discussed further below.

Finally, for the R-monopole we have the following relevant flux components (remember that $A_1=0$)
\begin{equation}
\begin{aligned}
\F^{z23}&=-\dt_1 H, \\
\F_1{}^{3z}&=\dt_2 H,\\
\F_1{}^{2z}&=-\dt_3 H,
\end{aligned}
\end{equation}
which gives the following integral
\begin{equation}
\begin{aligned}
4\p\m^R&=\int \F^{z23}d\tx_z\wedge d\tx_2 \wedge d\tx_3+ \int \F_1{}^{3z} dx^1 d\tx_3 d \tx_z+\int \F_1{}^{2z} dx^1 d\tx_2 d \tx_z\\
&=\int \F^{z23}dz\wedge dy^2 \wedge dy^3+ \int \F_1{}^{3z} dy^1 dy^3 d z+\int \F_1{}^{2z} dy^1 dy^2 d z\\
&=\int -\dt_1 H dz \ww dy^2 \ww dy^3 + \dt_2 H dy^1 \ww dy^3\ww dz-\dt_3 H dy^1 \ww dy^2 \ww dz\\
&=\int \big(\hr\dt_1H d\hr -\hr \dt_{\hr} H dy^1 \big)dzd\q=8\p^2 R_z h,
\end{aligned}
\end{equation}
where we used $\hr d\hr+y^1dy^1=0$ in the last line. Recall that in the case of the R-monopole a convenient parametrisation of the solution reads
\begin{equation}
\begin{aligned}
y^1&=y^1,\\
y^2&=\hr \cos{\q},\\
y^3&=\hr \sin{\q},\\
\end{aligned}
\end{equation}
and we use another notation for $\hr$ to avoid confusion between coordinate choices.

The net result is that the magnetic charge defined by~\eqref{magn-charge} is given by $\m=2\p R_zh$ which does not depend on the solution chosen, i.e. is T-duality invariant. The magnetic charge is equal to the magnetic charge $Q$ of the unsmeared NS5-brane solution.

\subsection{Noether current}

Invariance of the DFT action under generalised diffeomorphisms implies the existence of conserved charges from boundary integrals. The corresponding Noether procedure has been worked out in~\cite{Blair:2015eba,Park:2015bza} and the resulting conserved charge was found to give the mass of various solutions after reduction to $d=10$, as expected. A notable observation was made in~\cite{Park:2015bza}, that there is no contribution to the charge of the $5_2^2$-brane from the bulk part of the DFT action. A boundary term in the action is required in order to obtain nonvanishing value of the time component of the Noether current, $J^0$.

Following \cite{Berman:2011kg,Park:2015bza} we consider such a boundary contribution to the Hull--Hohm--Zwiebach action that turns it into a full Gibbons--Hawking type action upon section condition:
\begin{equation}
\begin{aligned}
S_B&=\int \dt_M \big(e^{-2d}K^M\big), \\
K^M&=4\mH^{MN}\dt_N d-\dt_N \mH^{MN}.
\end{aligned}
\end{equation}
The contribution to the Noether current from this term reads
\begin{equation}
J^M=\dt_N(2e^{-2d}K^{[M}\x^{N]})+e^{-2d}K_N \dt^M \x^N.
\end{equation}
Assuming that the generalised Killing vector $\xi^M$ is constant we find the time component of the current for the monopole solutions constructed in the section~\ref{DFTmon}:
\begin{equation}
\label{J0}
J^0 = \xi^0 H^{-2} \left( \partial_i H \partial_i H - H \partial_i \partial_i H\right) = -\xi^0 \partial_i \left(H^{-1}\partial_i H\right),
\end{equation}
where summation over $i \in \{1,2,3\}$ is done with a Kronecker delta, and $H$ is defined in~\eqref{H}. This result is the same for the H, KK, Q, and R-monopoles. In fact, $K^0 = 0$ for these solutions, which together with $\x^M = \mathrm{const}$ implies 
\begin{equation}
J^0 = \xi^0 \dt_M(e^{-2d}K^M),
\end{equation}
coincident with the boundary term in the Lagrangian that we started with.  In order to compute the corresponding conserved charge one may act in the way similar to the conventional electrodynamics
\begin{equation}
\label{charge0}
\begin{aligned}
   0&=\int_\mathbb{V} \dt_M J^M = \int_\mathbb{\partial\mathbb{V}} J^0 = Q(t_f)-Q(t_i),\\
Q(t)&=\int_{\mathbb{S}_t}J^0 d\mathbb{S}, 
\end{aligned}
\end{equation}
where $\mathbb{V}$ is the full $(10+10)$-dimensional doubled space, while $\mathbb{S}_t$ is its 19-dimensional slice at constant time $t$. This formally shows that the charge $Q(t)$ is the same at any two moments $t_f$ and $t_i$ and hence it is conserved, however, the procedure itself may not be well-defined. The subtlety is in the naive use of the Stokes' theorem, which in the case of extended space needs additional justification. As has been shown in  \cite{Naseer:2015fba} one may employ the naive generalization of the Stokes theorem and then make use of the section condition, that constrains the normal vector to the boundary. However, it is still not clear how to overcome the issue that the integration surface extends independently along both the dual and the ordinary coordinates, not to mention the issue of the dual time.

Defining the integration correctly is especially nontrivial for non-geometric solutions such as Q and R-monopoles, because in those cases some of the coordinates that the harmonic function $H(y^1, y^2, y^3)$ depends on are unphysical. Unless we integrate over the complete 19-dimensional space, it would be desirable to have integration over the physical space only, which for the Q-monopole is given by~\eqref{Q-phys}. Restricting the integration to a subspace of physical variables can be naturally done by inserting a delta-function in the integral. That is, instead of~\eqref{charge0} one would define
\begin{equation}
\label{charge1}
Q = \int dx_{\mathrm{phys}} dx_{\mathrm{dual}} \,J^0(x_{\mathrm{phys}}, x_{\mathrm{dual}}) \delta(x_{\mathrm{dual}}) = \int dx_{\mathrm{phys}} \, J^0(x_{\mathrm{phys}}, 0).
\end{equation}
Alternatively, one may try defining a dual coordinate dependent charge by
\begin{equation}
\label{charge2}
Q(x_{\mathrm{dual}}) = \int dx_{\mathrm{phys}} \,J^0(x_{\mathrm{phys}}, x_{\mathrm{dual}}),
\end{equation}
which is arguably an acceptable property for a solution that itself depends on a dual coordinate~\eqref{Q}. However, both definitions~\eqref{charge1},~\eqref{charge2} fail when applied to the Q or R-monopole as the corresponding integrals diverge for $J^0$ given by~\eqref{J0}.

As noted earlier, for the exotic monopole solutions~\eqref{J0} gives both the current component $J^0$ and the boundary term of the Lagrangian. This allows to directly compute the contribution of the DFT monopole into the full action, given that the bulk part of the action vanishes. Hence, we have
\begin{equation}\label{integral}
\begin{aligned}
S_B&=\int_\mathbb{V} \dt_M\big(e^{-2d}K^M\big) = \int_\mathbb{V}\frac{1}{H^2} \big( \partial_i H \partial_i H - H \partial_i \partial_i H\big)\\
&=k\int d^3y\, \dt_i\left(H^{-1}\dt_i H\right) = k\int_{\mathbb{S}_\infty^2}H^{-1}\dt_i H d^2\S^i  = 4\p k \fr{h}{1+\fr{h}{R}}\Big|_{R\to \infty}\\
&=4\p k h,
\end{aligned}
\end{equation}
where $\mathbb{S}_\infty^2$ is a sphere of the infinite radius $R\to \infty$ defined as $(y^1)^2+(y^2)^2+(y^3)^2=R^2$. The constant $k$ is related to the volume and for different solutions is given by
\begin{equation}
\begin{aligned}
\mbox{H and KK}:&&  &k = \mathrm{Vol}(\tx_1,\tx_2,\tx_3),\\
\mbox{Q}:       &&  &k = \mathrm{Vol}(\tx_1,\tx_2,x_3),\\
\mbox{R}:       &&  &k = \mathrm{Vol}(\tx_1,x_2,x_3).
\end{aligned}
\end{equation}
This comes from the integration over the coordinates that the harmonic function does not depend on. Such a volume pre-factor is to be expected, as similar contributions come from the HHZ action when the section condition is imposed. Note that some of the integration variables in~\eqref{integral} are actually dual coordinates in DFT and integration over these ensures finiteness of the result. 

On the other hand, for the smeared Q-monopole (the $5^2_2$-brane) one uses the harmonic function $H=1+h\log \m/\r$ with some cutoff $\m$~\eqref{log}, which yields zero after integration:
\begin{equation}
S_B = \left. \fr{2\pi h}{1+h\log \fr{\m}{\r}}\right|_{\r \to \infty} = 0.
\end{equation}
The same result has been obtained in \cite{Park:2015bza} where the regularization procedure of \cite{deBoer:2012ma} was then employed in order to obtain finite expression for the ADM mass. In contrast we do not recover this result for the full computation as the external directions $\{x^i,\tx_i\}$ are not toroidal.

In principle, it would be interesting to consider toroidal external space but keeping the dependence on the winding coordinates (localisation) and compute the charge. However, this falls beyond the scope of the present paper.

\noindent
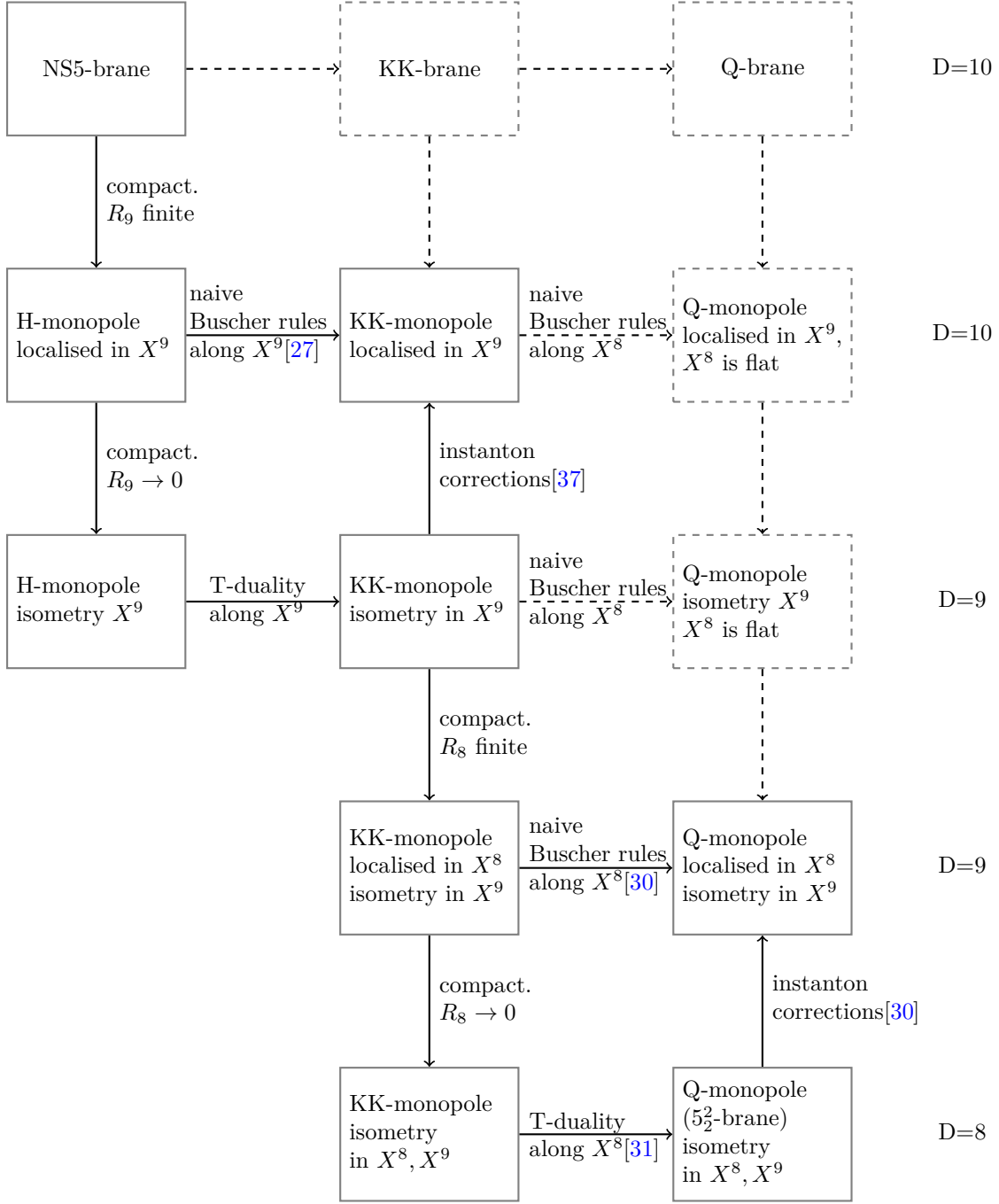
\begin{figure}[htbp]
\begin{tikzpicture}[thick,
                    text height=1ex,
                    text depth=.15ex,
                    scale=1
                    ]
\node at (-14,0) [format] (NS5) {\parbox{2.4cm}{\centering NS5-brane}};   
\node at (-14,-4) [format] (Hloc) {\parbox{2.4cm}{H-monopole \\ localised in $X^9$}};
\node at (-14,-8) [format] (H) {\parbox{2.4cm}{H-monopole \\ isometry $X^9$}};
\node at (-9,-0) [format1] (KKbr) {\parbox{2.4cm}{\centering KK-brane}};   
\node at (-9,-4) [format] (KKloc) {\parbox{2.4cm}{ KK-monopole\\ localised in $X^9$}};
\node at (-9,-8) [format] (KK) {\parbox{2.4cm}{KK-monopole \\ isometry in $X^9$}};
\node at (-9,-12) [format] (KK8) {\parbox{2.4cm}{KK-monopole \\ localised in $X^8$ \\ isometry in 
$X^9$}};
\node at (-9,-16) [format] (KK89) {\parbox{2.4cm}{KK-monopole \\ isometry \\ in $X^8,X^9$}};
\node at (-4,-0) [format1] (Qbr) {\parbox{2.4cm}{\centering Q-brane}};   
\node at (-4,-4) [format1] (Qhloc) {\parbox{2.4cm}{Q-monopole \\ localised in $X^9$, $X^8$ is flat}};
\node at (-4,-8) [format1] (Q) {\parbox{2.4cm}{Q-monopole \\ isometry 
$X^9$ \\ $X^8$ is flat }};
\node at (-4,-12) [format] (Q8) {\parbox{2.4cm}{Q-monopole \\ localised in $X^8$ \\ isometry in 
$X^9$}};
\node at (-4,-16) [format] (522) {\parbox{2.4cm}{Q-monopole \\ ($5^2_2$-brane) \\ isometry \\ in 
$X^8,X^9$}};
\node at (-1,-0) [format2] (1) {D=10};   
\node at (-1,-4) [format2] (2) {D=10};
\node at (-1,-8) [format2] (3) {D=9};
\node at (-1,-12) [format2] (4) {D=9};
\node at (-1,-16) [format2] (4) {D=8};
\node at (-14,-2) [format2,right=0] {\parbox{1.4cm}{compact.\\ $R_9$ finite}};
\node at (-14,-6) [format2,right=0] {\parbox{1.4cm}{compact.\\ $R_9 \to 0$}};
\node at (-11.5,-4) [format2,above=-0.1] {\parbox{2.2cm}{naive \\ Buscher rules\\ along $X^9$\cite{Jensen:2011jna}}};
\node at (-11.5,-8) [format2] {\parbox{1.6cm}{T-duality\\ along  $X^9$}};
\node at (-9,-6) [format2,right=0] {\parbox{1.8cm}{instanton\\ corrections\cite{Gregory:1997te}}};
\node at (-9,-10) [format2,right=0] {\parbox{1.4cm}{compact.\\ $R_8$ finite}};
\node at (-9,-14) [format2,right=0] {\parbox{1.4cm}{compact.\\ $R_8 \to 0$}};
\node at (-4,-14) [format2,right=0] {\parbox{1.8cm}{instanton\\ corrections\cite{Kimura:2013zva}}};
\node at (-6.5,-16) [format2,above=-0.31] {\parbox{2cm}{T-duality\\ along  $X^8$\cite{deBoer:2012ma}}};
\node at (-6.4,-12) [format2,above=-0.1] {\parbox{2.2cm}{naive \\ Buscher rules\\ along $X^8$\cite{Kimura:2013zva}}};
\node at (-6.4,-4) [format2,above=-0.1] {\parbox{2.2cm}{naive \\ Buscher rules\\ along $X^8$}};
\node at (-6.4,-8) [format2,above=-0.1] {\parbox{2.2cm}{naive \\ Buscher rules\\ along $X^8$}};

\path[->] (NS5) edge   (Hloc);
\path[->] (Hloc) edge   (H);
\path[dashed,->] (KKbr) edge   (KKloc);
\path[<-] (KKloc) edge   (KK);
\path[->] (KK) edge   (KK8);
\path[->] (KK8) edge   (KK89);
\path[<-] (Q8) edge   (522);
\path[->] (H) edge   (KK);
\path[->] (Hloc) edge   (KKloc);
\path[->] (KK89) edge   (522);
\path[->] (KK8) edge   (Q8);
\path[dashed,->] (KK) edge   (Q);
\path[dashed,->] (Qbr) edge   (Qhloc);
\path[dashed,->] (Qhloc) edge   (Q);
\path[dashed,->] (Q) edge   (Q8);
\path[dashed,->] (NS5) edge   (KKbr);
\path[dashed,->] (KKbr) edge   (Qbr);
\path[dashed,->] (KKloc) edge   (Qhloc);
\end{tikzpicture}
\caption{\sl Systematics of backgrounds with H, geometric $\t$ and Q fluxes and their relations. Note that in \cite{Kimura:2013zva} Q-monopole localised in both $X^8$ and $X^9$ has been constructed with both these directions being compact. \label{fig:systematics}}
\label{figBG}
\end{figure}

\section{Discussion and speculations}

The DFT-monopole solution of \cite{Berman:2014jsa} is one of the first explicit solutions of the DFT equations of motion. This solution was known to reproduce the background of NS5-brane and KK-monopole depending on the identification of the physical subset of coordinates among 10+10 coordinates of DFT. The harmonic function of this solution depends only on 3 coordinates (or in the localised case, 4), which are a priori  not identified neither with physical nor with winding coordinates.

In this work we have shown that certain choices of physical coordinates give rise to backgrounds, which can be understood as Q- and R-monopoles. As a source of Q-flux we obtain a background whose harmonic function depends on one winding coordinate while R-flux corresponds to a harmonic function that depends on two winding coordinates. As an explicit check we have shown that smearing the Q-monopole along the winding coordinate gives just the known $5^2_2$-brane of \cite{deBoer:2012ma}. Interestingly, the Q-monopole solution does not have nontrivial monodromies and shows its non-geometric nature only via dependence on winding coordinates. Alternatively, in the case of the $5^2_2$-brane the non-geometry manifests itself via non-trivial monodromy properties when going around the (smeared) brane.

Dependence of supergravity fields on winding coordinates has been discussed before in the literature \cite{Harvey:2005ab,Jensen:2011jna,Kimura:2013zva}. This effect is due to taking into account worldsheet instantons of the $(1+1)$-dimensional sigma model understood as a gauge theory. These instanton corrections modify the backgrounds of the H-monopole, KK-monopole and $5^2_2$-brane by contributing to the 4-point interactions in such a way that the form of the background remains the same, but the harmonic function acquires dependence on a winding coordinate. It is important to mention that although our harmonic function does have dependence on a winding coordinate, say $\tx_1$, this does not break the section condition of DFT as the function does not depend on the dual coordinate $x^1$ at the same time. Another point to be mentioned here is that the coordinate added by instanton corrections is circular and the corresponding harmonic function includes hyperbolic cosines and sines or Bessel functions. In our case these coordinates in principle can be kept noncompact and the harmonic function is just some power of the distance $r$. However, this returns us back to the discussion on whether DFT with section condition imposed lives on a torus or on a space of general topology. In any case, the solutions presented above are valid up to an appropriate choice of the harmonic function (smearing). With all these reservations we may present a diagram showing systematics of the backgrounds with H, geometric $\t$ and Q fluxes from the point of view of DFT, see figure~\ref{fig:systematics}. Although the R-monopole background can be trivially included into the diagram, we intentionally keep it off the picture as it does not connect to any known node as $5^2_2$-brane, for example.

As a check that the obtained backgrounds indeed source the desired fluxes we have computed them explicitly using the notion of the DFT generalised flux. Indeed, we show that H-, KK-, Q- and R-monopole source H-, $\t$, Q- and R-fluxes respectively, which satisfy generalised Bianchi identities. The computed components of the generalised flux allow us to consider the notion of the magnetic charge for these solutions. Following \cite{Blair:2015eba} this is defined to be
\begin{equation}
4\p \m=\int_\S \F_{MNK}d\XX^M \wedge d\XX^N \wedge d\XX^K,
\end{equation}
and explicit calculation shows that $\m=2\p R_z h=Q$, where $R_z$ is the radius of the compactified $z$ direction and $Q$ is the magnetic charge of the full (unsmeared) NS5-brane. Since $\m$ is the same for all solutions, one may call it the magnetic charge of DFT monopole.

One should note here, that the generalised flux for KK- and Q-monopoles has not only the expected components of the form $\t^z{}_{12}{}$ or $Q_z{}^{12}$, but also components with two indices equal $\t^a{}_{ab}$ and $Q_a{}^{ab}$ (no sum) as well as components of the gaugings $\F_M$. Although these are not welcome in the models of conventional supergravity compactifications, they in principle can be present in the $10$-dimensional theory. However, given the definition of the magnetic charge above, they do not contribute to $\m$.

Although certain properties of Q- and R-monopole have been revealed, the present work can not be viewed as an exhaustive study of non-geometric branes. There still are many open questions left. Firstly, one may be interested to look at the equations of motion of conventional supergravity and to what extent the presented solutions solve them. This may give a hint to the meaning of the dual coordinates inside the harmonic function. 

From the DFT point of view it is interesting to see how many supersymmetries these backgrounds preserve and whether the resulting Killing spinors depend on dual coordinates. We expect these backgrounds to preserve half of the maximal supersymmetry in analogy with the $5^2_2$-brane background. Also one may study non-commutativity and non-associativity of the doubled NLSM on such backgrounds in the spirit of \cite{Blair:2014kla}.

The most obvious extension of the presented work is to consider the localised version of the DFT-monopole of \cite{Berman:2014jsa}. Following the same logic one considers other choices of the set of physical coordinates inside the doubled space to get
\begin{equation}
\begin{aligned} 
x^\m&=(z,y^1,y^2,\ty_3,x^r), && \mbox{localised KK-monopole}\\ 
x^\m&=(z,y^1,\ty_2,\ty_3,x^r), && \mbox{localised Q-monopole}\\
x^\m&=(\tz,\ty_1,\ty_2,\ty_3,x^r), && \mbox{constant background}\\
x^\m&=(z,\ty_1,\ty_2,\ty_3,x^r), && \mbox{localised R-monopole}.
\end{aligned}
\end{equation}
Also here one may look into the monopole solutions of exceptional field theory (EFT) which encode the M5-brane according to \cite{Berman:2014jsa}. Other choices of the physical subset may reproduce the exotic branes of M-theory classified in \cite{deBoer:2012ma}.

Finally, it would be interesting to investigate near-horizon limit of the Q- and R-monopole in the same way as the near-horizon limit of NS5-brane is studied \cite{Aharony:1998ub}.

\section*{Acknowledgements}
ETM thanks David Berman and Chris Blair for valuable comments on this paper. IB is grateful for the hospitality of the Albert-Einstein-Institut of MPI Potsdam, where part of this work was done. The work of IB was supported by the Russian Government programme of competitive growth of Kazan Federal University. The work of ETM was supported by the Alexander von Humboldt Foundation and in part by the Russian Government programme of competitive growth of Kazan Federal University.

\appendix

\section{Notation and conventions}
\label{conventions}

The notation for indices used in this paper is as follows
\begin{equation}
\begin{aligned}
& M,N,K \ldots =1,\ldots 20, && \mbox{curved indices for the doubled space};\\
& A,B,C \ldots =1,\ldots 20, && \mbox{flat indices for the doubled space};\\
& \m,\n,\r,\s, \ldots =0,\ldots 9, && \mbox{space-time curved indices};\\
& k,l,m,n \ldots =0,\ldots 9, && \mbox{space-time flat indices};\\
& i,j =1,2,3 && \mbox{indices for transverse coordinates $x^i$};\\
& \ba,\bb,\bc, \ldots = z,1,2,3 && \mbox{flat indices for transverse directions $x^z$ and $x^{1,2,3}$};\\
& r =0,5,6,7,8,9 && \mbox{index for transverse coordinates $x^r$};\\
& \bz,\bar{1}, \bar{3} && \mbox{flat indices for directions 1,3 and $z$};\\
& \a,\b  && \mbox{run 1,2 for Q-monopole and 2,3 for R-monopole labelling};\\
&         && \mbox{some of the transverse coordinates}\\
& \bar{\a},\bar{\b}  && \mbox{the same as above, but flat};
\end{aligned}
\end{equation}

The $O(d,d)$ invariant metric $\h_{MN}$ and the flat generalised metric are defined as
\begin{equation}
\begin{aligned}
\h_{MN}=
\begin{bmatrix}
0 & 1 \\
1 & 0
\end{bmatrix}, && 
\mH_{AB}=
\begin{bmatrix}
\d^a_b & 0 \\
0 & \d^b_a
\end{bmatrix}.
\end{aligned}
\end{equation}
The generalised Lie derivative of DFT and the C-bracket have the usual form
\begin{equation}
\begin{aligned}
[V_1,V_2]_C^M&=\fr12(\mc{L}_{V_1}V_2-\mc{L}_{V_2}V_1)^M,\\
\mc{L}_{V_1}V_2^M&=2V_{[1}^N\dt_N V_{2]}^M+\h^{MN}\h_{KL}\dt_N V_{1}^KV_{2}^L.
\end{aligned}
\end{equation}

\section{Duality invariant formulation of fluxes}
\label{torsion}

The conventional geometric $\t$-flux is defined via Maurer-Cartan forms as a torsion
\begin{equation}
de^{m}=-\fr12 \t^{m}{}_{nk}e^{n}\wedge e^{k},
\end{equation}
where Latin indices are used for flat directions and $e^{m}=e^{m}_\m dx^\m$ is a 1-form that defines the vielbein. This equation can be written in equivalent form by making use of the Lie bracket of two vector fields
\begin{equation}
\label{bracket_e}
[e_{m},e_{n}]=f^{k}{}_{mn}e_{k}.
\end{equation}
Here the inverse vielbein is a vector field $e_{m}=e_{m}^\m\,\dt_\m$ and $f^{m}{}_{nk}=2e^{m}_{\m}e^\n_{[n}\dt_\n e^\m_{k]}$.

It was suggested in \cite{Grana:2008yw} to generalise the construction \eqref{bracket_e} to the case of Double Field Theory using the C-bracket $[,]_C$, which is a natural multiplication of generalised vectors  in Double Field Theory
\begin{equation}
\label{bracket_E}
[E_{A},E_{B}]^M_C=\F^{C}{}_{AB}E^M_{C},
\end{equation} 
where $E_B^{M}$ is a generalised vielbein defined as
\begin{equation}
\mH^{MN}=E^M_{A}E^N_{B}\mH^{AB}.
\end{equation}
The diagonal form of the flat generalised metric $\mc{H}_{AB}=\mbox{diag}[h_{mn},h^{mn}]$ corresponds to the two natural gauge choices for the generalised vielbein, called B- and $\b$-frame. Each of these corresponds to either non-zero Kalb--Ramond field $B$ or the bivector $\b$. Although having both $B$ and $\b$ non-vanishing is inconsistent with the counting of degrees of freedom, it is convenient for calculational purposes to write the generalised vielbein in a $(B,\b)$-frame
\begin{equation}
\label{vielbein}
\begin{aligned}
	& E^M_A=
				\begin{bmatrix}
					e^\m_{m} & & -e^{n}_\r\b^{\r\m}\\
						&\\
					-e_{m}^\r B_{\r\n}  & & e^n_\n+e^n_\r\b^{\r\s}B_{\s\n}
				\end{bmatrix},  &&
	 E_M^A=
			\begin{bmatrix}
				e_\m^{m}+e_\r^{m}\b^{\r\s}B_{\s\m} & & -e^{m}_\r\b^{\r\n}\\
					&\\
				-e_{n}^\r B_{\r\n} & & e_n^\n
			\end{bmatrix}.			
\end{aligned}
\end{equation}

One can think of the generalised vielbein as Scherk--Schwarz twist matrices \cite{Grana:2012rr} and the structure constants $\F^{A}{}_{BC}$ are thus gaugings of the corresponding supergravity \cite{Geissbuhler:2011mx}. Given the definition of the C-bracket, the generalised flux and the additional gauging $\F_A$ can be written as
\begin{equation}
\begin{aligned}
\F^{A}{}_{BC}&=2E^{A}_M E_{[B}^N\dt_N E_{C]}^M-E^{A}_M \h^{MN}\h_{KL}\dt_N E_{[B}^K E_{C]}^L,\\
\F_A&=\dt_M E^M_A+2E^M_A\dt_Md.
\end{aligned}
\end{equation}

Using the notation of \cite{Geissbuhler:2013uka} flux components in terms of the fundamental fields can be written as
\begin{equation}
\label{nf}
\begin{aligned}
{\F}_{mnk} =&\ 3\left[ \nabla_{[m} B_{nk]}  - B_{l[m} \tilde\nabla^l B_{nk]} \right]\, ,\\
{\F}_{mn}{}^{k} =&\ 2 \G_{[mn]}{}^k + \tilde \nabla^k B_{mn} + 2 \G^{lk}{}_{[m} B_{n]l} + \b^{kl}{\F}_{lmn}\, ,\\
{\F}_k{}^{mn} =&\ 2 \G^{[ab]}{}_k +\partial_k\b^{mn}+  B_{kl} \tilde \partial^l\b^{mn} + 2 {\F}_{lk}{}^{[m}\b^{n]l}-{\F}_{lpk} \beta^{lm}\beta^{pn}\, ,\\
{\F}^{mnk} =&\ 3\left[-\b^{l[m}\nabla_l \b^{nk]}
+ \tilde \nabla^{[m} \b^{nk]} +B_{lp} \tilde \nabla^p\b^{[mn} \b^{k]l} - \b^{l[m}\b^{n|p|} \tilde\nabla^{k]}B_{lp}
 \right ]\\
 & +\b^{ml}\b^{ np}\b^{kq}{\F}_{lpq},\\ 
\F_{m}=&-\tilde{\nabla}^n B_{mn} - \G^{kl}{}_m B_{kl} - \G_{km}{}^{k} + 2 B_{mk}\tilde{\nabla}^k d + 2\nabla_{m}d\, ,\\
\F^a=&-\G^{km}{}_{k}+\tilde{\nabla}^k\b^{ml}B_{kl}-\G^{lm}{}_n\b^{nk}B_{kl}-\b^{mk}\tilde{\nabla}^lB_{kl}+ 2\tilde{\nabla}^m d +2\b^{mk}B_{kl}\tilde{\nabla}^l d\\
&  +2\b^{mn}\nabla_n d -\nabla_n\b^{mn} - \G_{kl}{}^m\beta^{kl}.
\end{aligned} 
\end{equation}
Here the following conventions have been adopted
\begin{equation}
\begin{aligned}
 B_{mn} &= e_m{}^\m e_n{}^\n B_{\m\n}, & \dt_m &= e_m{}^\m \dt_\m, \\
 \b^{mn} &= e^m{}_\m e^n{}_\n  \b^{\m\n}, & \tdt^m &= e^m{}_\m \tdt^\m.
\end{aligned}
\end{equation}
The covariant derivatives are defined in the same way as in \cite{Geissbuhler:2013uka}
\begin{equation}
\begin{aligned}
\nabla_m B_{nk}&=\dt_m B_{nk}+2\G_{m[n}{}^lB_{k]l}, &
\tilde \nabla^m B_{nk}&=\tdt^mB_{nk}-2\G^{ml}{}_{[n}B_{k]l},\\
\nabla_m \b^{bc}&=\dt_m\b^{nk}-2\G_{ml}{}^{[n}\beta^{k]l}, &
\tilde \nabla^m \b^{nk}&=\tdt^m\b^{nk}+2\G^{m[n}{}_{l}\beta^{k]l}.
\end{aligned} 
\end{equation}
With $\G$-symbols given by the following expressions
\begin{equation}
\label{Gammas}
\begin{aligned}
\G_{mn}{}^k &= e_m{}^\m \dt_\m e_n{}^\n e^k{}_\n, \\
\G^{mn}{}_k &= e^m{}_\m \tdt^\m e^n{}_\n e_k{}^\n.
\end{aligned}
\end{equation}

Using the notations  ${H},{F},{Q}$ and ${R}$ for the corresponding components of the generalised flux one may write the equation \eqref{bracket_E} as follows
\begin{equation}
\begin{aligned}
&[E_{m},E_{n}]={F}^{k}{}_{mn}E_{k}+{H}_{mnk}E^{k},\\
&[E^{m},E_{n}]={F}^{m}{}_{nk}E^{k}+{Q}^{km}{}_{n}E_{k},\\
&[E^{m},E^{n}]={Q}^{mn}{}_{k}E^{k}+{R}^{mnk}E_{k}.
\end{aligned}
\end{equation}
One should note, however, that the generalised vielbein used above is written in the $(B,\b)$-frame and hence contains $1/2d(d-1)$ more degrees of freedom than usual. For this reason the above expression should be understood just as a convenient tool to incorporate both B- and $\b$-frames in a single expression. Before going to calculations one should choose the frame to work in, taking into account that H-flux identically vanishes in the $\b$-frame, while R-flux vanishes in the B-frame.

\section{Bianchi identities}
\label{bianchi}

It has been shown in \cite{Geissbuhler:2013uka} that the equations of motion of DFT and consistency of the algebra impose certain conditions on the generalised flux which can be written in the form of Bianchi identities on its components. Here we copy them from that work without any derivation just to present the input used in the main text. Hence, the generalised Bianchi identities come from the requirement that the generalised flux (with flat indices) transform as a scalar under generalised diffeomorphisms:
\begin{equation}
\label{obi}
\begin{aligned}
\DD_{[m}H_{nkl]} -\frac{3}{2}H_{p[mn} \tau_{kl]}{}^p &=0 ,\\
3\DD_{[m}\tau_{nk]}{}^l - \DD^d H_{mnk} +3\tau_{[mn}{}^p\tau_{k]p}{}^l
-3Q_{[m}{}^{lp}H_{nk]p} &= 0 ,\\
2\DD_{[m}Q_{n]}{}^{kl} +2 \DD^{[k}\tau_{mn}{}^{l]}
- \tau_{mn}{}^p Q_p{}^{kl} - H_{mnp}R^{pkl} + 4Q_{[m}{}^{p[k}\tau_{n]p}{}^{l]} &=0 ,\\
3\DD^{[m}Q_l{}^{nk]} - \DD_l R^{mnk} +3Q_p{}^{[mn}Q_l{}^{k]p}
-3\tau_{lp}{}^{[m}R^{nk]p} &= 0 ,\\
\DD^{[m}R^{nkl]} -\frac{3}{2}R^{p[mn} Q_p{}^{kl]} &=0;
\end{aligned}
\end{equation}
the requirement that the DFT action transform as a scalar under local $O(d)\times O(d)$ transformations:
\begin{equation}
\begin{aligned}
\DD^k H_{mnk} +\DD_k\tau_{mn}{}^k +2\DD_{[m}\F_{n]}
-\F^kH_{mnk} - \F_k\tau_{mn}{}^k&=&0,\\
\DD^k \tau_{cm}{}^n +\DD_k Q_m{}^{nk} +\DD_{m}\F^{n} - \DD^{n}\F_{m}
-\F^k \tau_{cm}{}^n - \F_k Q_m{}^{nk}&=&0 ,\\
\DD_k R^{mnk} +\DD^k Q_k{}^{mn} +2\DD^{[m}\F^{n]}
-\F_k R^{mnk} - \F^k Q_k{}^{mn}&=&0;
\end{aligned}
\end{equation}
and the strong constraint:
\begin{equation}
\DD^m\F_m + \DD_m\F^m - \F^m\F_m +\frac{1}{6}H_{mnk}R^{mnk}
+ \frac{1}{2} \tau_{mn}{}^c Q_c{}^{mn} =0 .
\end{equation}
Note that all indices here are flat, and $\DD_A=E_A^M\dt_M$, which is different from just $(\dt^m,\tdt_m)$ in the previous section.

\bibliographystyle{utphys}
\bibliography{bib}
\end{document}